\documentclass[aps,prl,12pt,twocolumn,tightenlines,superscriptaddress,
amsfonts,amssymb,amsmath,byrevtex,showpacs,nofootinbib]{revtex4}
\usepackage{graphics}
\usepackage{epsfig,subfigure}

\newcounter{mnotecount}[section]
\renewcommand{\themnotecount}{\thesection.\arabic{mnotecount}}
\newcommand{\mnote}[1]
{\protect{\stepcounter{mnotecount}}$^{\mbox{\footnotesize  $
      \bullet$\themnotecount}}$ \marginpar{\raggedright\tiny
    $\!\!\!\!\!\!\,\bullet$\themnotecount: #1} }

\begin{document}
\newcommand{\dC}{\mathbb C}
\newcommand{\dR}{\mathbb R}
\newcommand{\dQ}{\mathbb Q}
\newcommand{\dZ}{\mathbb Z}
\newcommand{\dN}{\mathbb N}
\newcommand{\id}{\mathbb I}

\author{Przemys{\l}aw Ma{\l}kiewicz}
\email{pmalk@fuw.edu.pl}
\author{W{\l}odzimierz Piechocki}
\email{piech@fuw.edu.pl} \affiliation{  Theoretical Physics
Department, Institute for Nuclear Studies,
\\ Ho\.{z}a 69, 00-681 Warszawa, Poland}

\date{\today}

\title{Foamy structure of spacetime}

\begin{abstract}
We examine  spectrum of the  physical volume operator within the
non-standard loop quantum cosmology. The spectrum is discrete with
equally distant levels defining  a quantum of the volume. The
discreteness may imply a foamy structure of spacetime at
semi-classical level which may be detected in astro-cosmo
observations.
\end{abstract}
 \maketitle

\noindent{\it 1. Introduction.} Various forms of  discreteness of
spacetime underly many approaches in fundamental physics. Just to
name a few: noncommutative geometry \cite{Heller:2006uq}, causal
sets approach \cite{Rideout:2008df}, gravitational Wilson loops
\cite{Hamber:2009uz}, Regge calculus \cite{Bahr:2009qc}, path
integral over geometries \cite{Ambjorn:2009ts}, spin foam model
\cite{Kaminski:2009fm}, and categories \cite{Baez:2009as}. The
discreteness may translate at the semi-classical level into a
foamy structure of space. Such expected property of spacetime
creates large activity in observational astrophysics and cosmology
(see, e.g. Lorentz and CPT violation \cite{Kostelecky:2008ts},
dispersion of cosmic photons \cite{AmelinoCamelia:2009pg},
electrons \cite{Galaverni:2008yj} and neutrinos
\cite{Ellis:2008fc}, birefringence effects \cite{Gleiser:2001rm}).

The goal of our paper is presentation of the physics of geometry
at short distances. We study the spectrum of the volume operator.
We find that the spectrum is bounded from below and discrete. The
minimum distance between the levels of the spectrum defines a {\it
quantum} of the volume.  We suggest  that there may exist
elementary {\it quanta} of energy connected with the multiplicity
of the quantum of the volume, i.e. with a {\it foamy} like
structure of space. They may have a form of elementary objects
like photon, electron, proton, dark matter particle, etc.

Our results suggest that the foamy structure of space is likely to
be a real property of the Universe so its identification via
astro-cosmo observations has sound motivation and is important for
the fundamental physics.

Our results are obtained within the non-standard loop quantum
cosmology (LQC) developed recently
\cite{Dzierzak:2008dy,Malkiewicz:2009zd,Dzierzak:2009ip,Dzierzak:2009dj,Malkiewicz:2009qv}.
In this method one first solves classical constraints to identify
the  physical phase space and  finds an algebra of observables,
then one imposes quantum rules. Standard LQC
\cite{Ashtekar:2003hd,Bojowald:2006da} means first quantization of
the kinematics, then imposition of constraints in the form of
operators acting on the kinematical Hilbert space. Both methods
should `commute', i.e. give the same results. In the case of
quantization of the Maxwell electrodynamics such treatment of
constraints leads to equivalent results \cite{IBB}. Thus, another
aim of our paper is of methodological nature: testing the
equivalence of both methods in the case of gravitational
interaction.

This Letter is meant to address a wide physical community. It
popularizes and interprets the results of a `technical' paper
\cite{Malkiewicz:2009qv} directed to experts in quantum cosmology,
and submitted for publication elsewhere.

A direct way of testing  the singularity aspects of a given
cosmological model is by an examination of the energy density of
matter as a function of time \cite{Malkiewicz:2009zd}. However,
the geometry of space, as a function of time, is sensitive to
these aspects too.  We carry out the corresponding discussion in
\cite{Malkiewicz:2009qv}.

\noindent{\it 2. Modified Hamiltonian.} In what follows, for
simplicity of exposition, we restrict ourselves to the quantum
flat Friedmann-Robertson-Walker (FRW) model with massless scalar
field.

The classical dynamics may be defined by the FRW Hamiltonian
\begin{equation}\label{hham}
 H= N\bigg(-\frac{3}{8\pi G\gamma^2}\;\beta^2\;v + \frac{p_{\phi}^2}{2\,
  v}\bigg)\approx 0\texttt{} ,
\end{equation}
which is known to be a dynamical constraint
\cite{Malkiewicz:2009zd}; $N$ denotes the lapse function and
$\gamma$ is the so-called Barbaro-Immirzi parameter; $(\beta, v,
\phi,p_{\phi})$ are the kinematical phase space variables;
$v^{1/3} \sim a$, where $a$ is the scale factor; $\beta
\sim\dot{a}/a$ so it corresponds to the Hubble parameter;
$p_{\phi}$ is the momentum of the massless scalar field $\phi$.

The Hamiltonian modified by the so-called {\it holonomy} functions
specific to LQC, corresponding to (\ref{hham}), is found to be
\cite{Dzierzak:2009ip}
\begin{equation}\label{haam}
 H^{(\lambda)}= N\bigg(-\frac{3}{8\pi G \gamma^2}\;\frac{\sin^2(\lambda
\beta)}{\lambda^2}\;v + \frac{p_{\phi}^2}{2\, v}\bigg)\approx 0.
\end{equation}
The parameter $\lambda$ is a {\it free} parameter of the
non-standard LQC method parameterizing holonomies. It is clear
that in the limit $\lambda \rightarrow 0$ the Hamiltonian
(\ref{haam}) turns into (\ref{hham}). In what follows we consider
(\ref{haam}) in the gauge
\begin{equation}\label{ggg}
N^{-1}:=\frac{3}{8 \pi G \gamma^2 v} \;\Big(\kappa \gamma |p_\phi|
+ v\,\frac{|\sin(\lambda \beta)|}{\lambda}\Big),
\end{equation}
where $\kappa^2 \equiv 4\pi G/3$. Consequently, (\ref{haam}) leads
to the dynamical constraint
\begin{equation}\label{ham}
H^{(\lambda)}:= \kappa \gamma |p_\phi| - v\, \frac{|\sin(\lambda
\beta)|}{\lambda}\approx 0.
\end{equation}
The FRW constraint in the gauge corresponding to (\ref{ggg}),
reads
\begin{equation}\label{ham2}
H:= \kappa \gamma |p_\phi| - v\,  |\beta|\approx 0.
\end{equation}

Since $\sin(\cdot)$ is bounded from above, there exists $\epsilon
\in \dR$, due to (\ref{ham}), such that $v > \epsilon > 0$. Thus,
there exists $\varepsilon \in \dR$ such that the scale factor $a >
\varepsilon> 0$. As $\sin(\cdot)$ is a periodic function, the
variable $\beta$ which occurs in (\ref{ham}) is bounded. Thus, the
Hubble parameter is bounded, which means that there is no
Big-Bang. The variables $\beta$ and $v$ which satisfy (\ref{ham2})
do not  have such properties so there is Big-Bang. Thus, the {\it
modification} of the classical Hamiltonian turns Big-Bang into
Big-Bounce.  In what follows we show that {\it quantization} of
the bouncing dynamics inevitably leads to {\it discrete} spectrum
of the volume operator.

It turns out that the {\it physical} phase space may be
parameterized by the  elementary {\it observables}
\cite{Dzierzak:2009ip},
\begin{equation}\label{obb}
\mathcal{O}_1:= p_{\phi},~~~~\mathcal{O}_2:= \phi -
\frac{\textrm{sgn}(p_{\phi})}{3\kappa}\
\textrm{arth}\big(\cos(\lambda \beta)\big),
\end{equation}
i.e. functions  having vanishing Poisson bracket with (\ref{ham})
on the constraint surface $H^{(\lambda)}\approx 0$.

The classical dynamics has been solved analytically
\cite{Dzierzak:2009ip} and the explicit form of the solution for
the variable $v$, which is of interest in the present paper, is
given by
\begin{equation}\label{vol2}
    v(\phi) = \kappa\gamma\lambda\,
    |\mathcal{O}_1|\,\cosh3\kappa  (\phi-
    \mathcal{O}_2).
\end{equation}
The variable $\phi$ changes monotonically with an evolution so it
has been chosen to be an evolution parameter of the system
\cite{Dzierzak:2009ip}.

\noindent{\it 3. Volume Operator.} The variable $v$ has the
interpretation of a volume of some piece of space
\cite{Ashtekar:2006wn}. To define quantum operator corresponding
to $v$, we use the classical observables (\ref{obb})
\begin{equation}\label{vol}
v = |w|,~~~w :=
\kappa\gamma\lambda\;\mathcal{O}_1\;\cosh3\kappa(\phi-
\mathcal{O}_2).
\end{equation}
Thus, quantization of $v$ reduces to the  quantization problem of
$w$. Quantization of the latter may be done in a standard way as
follows \cite{Malkiewicz:2009qv}
\begin{eqnarray}\nonumber
&&\hat{w}\,f(x) :=
    \kappa\gamma\lambda\,\frac{1}{2}\,\big(
    \widehat{\mathcal{O}}_1\,\cosh3\kappa  (\phi-
    \widehat{\mathcal{O}}_2)\\ \label{c1}
     &+&\cosh3\kappa  (\phi-
    \widehat{\mathcal{O}}_2)\;\widehat{\mathcal{O}}_1\big) f(x),
\end{eqnarray}
where $f \in  L^2 (\dR)$, and where $\phi$ is a scalar field used
both at classical and quantum levels as an evolution parameter
\cite{Malkiewicz:2009zd}.

For the elementary observables $\mathcal{O}_1$ and $\mathcal{O}_2$
we use the Schr\"{o}dinger representation
\begin{eqnarray}\nonumber
\mathcal{O}_1 \longrightarrow \widehat{\mathcal{O}}_1 f(x)&:=&
-i\,\hbar\,\partial_x f(x),\\ \label{rep1} \mathcal{O}_2
\longrightarrow \widehat{\mathcal{O}}_2 f(x)&:=& \widehat{x} f(x)
:= x f(x).
\end{eqnarray}

In the representation (\ref{rep1}) an explicit form of the
operator $\hat{w}$ is
\begin{eqnarray}\nonumber
\hat{w}\,f(x)= i\,\frac{\kappa\gamma\lambda\hbar}{2}\big(
    2 \cosh3\kappa(\phi-x)\;\frac{d}{dx}\\ \label{repp1}
     -3\kappa\sinh3\kappa
    (\phi-x)\big)\,f(x).
\end{eqnarray}

\noindent{\it 4. Eigenvalue problem.} It turns out that the
solution to the eigenvalue problem
\begin{equation}\label{eq4}
\hat{w}\, f_a (x) = a\,f_a (x),~~~a \in \dR ,
\end{equation}
reads \cite{Malkiewicz:2009qv}
\begin{equation}\label{eq5}
f_a (x):= \frac{\sqrt{\frac{3\kappa}{\pi}}\exp\big(i \frac{2
a}{3\kappa^2 \gamma\lambda\hbar}\arctan
    e^{3\kappa(\phi-x)}\big)}{\cosh^{\frac{1}{2}}3\kappa(\phi-x)}.
\end{equation}
The condition $\,\langle f_b|f_a\rangle = 0\,$ leads to
\begin{equation}\label{eqq5}
    a-b=6\kappa^2\gamma\lambda\hbar\,m =8\pi
    G\gamma\lambda\hbar\,m,
\end{equation}
where $m\in \mathbb{Z}$. Thus, the set
\begin{equation}\label{set1}
\mathcal{F}_b:=\{~f_a\;|\; a = b + 8\pi G\gamma\lambda\hbar\, m\},
\end{equation}
where $b \in \dR$, is orthonormal. Each subspace $\mathcal{F}_b
\subset L^2(\dR)$ spans a pre-Hilbert space.  The completion of
each span $\mathcal{F}_b$ in the norm of $L^2(\dR)$ defines an
infinite dimensional  {\it separable} Hilbert space
$\mathcal{H}_b$. Since
\begin{equation}\label{set2}
\langle f_b|\hat{w}f_a\rangle - \langle \hat{w}f_b|f_a\rangle =
(a-b)\,\langle f_b|f_a\rangle,
\end{equation}
the operator $\hat{w}$ is symmetric on $\mathcal{F}_b$ for any $b
\in \dR$.  In fact, it is  a {\it self-adjoint} operator on the
span of $\mathcal{F}_b$ (see, \cite{Malkiewicz:2009qv} for a
proof).

\noindent{\it 5. Spectrum.} Due to the the relation (\ref{vol})
and the spectral theorem on self-adjoint operators \cite{DS,RS},
we may carry out quantization of the volume function on each
$\mathcal{F}_b$ as follows
\begin{equation}\label{sp1}
 v = |w|~~~\longrightarrow~~~\hat{v} f_a :=  |a| f_a .
\end{equation}

A common feature of all $\mathcal{F}_b$ is the existence of the
minimum gap $\bigtriangleup  := 8\pi G\gamma\hbar\,\lambda\;$ in
the spectrum, which defines a  {\it quantum} of the volume.  In
the limit $\lambda \rightarrow 0$, corresponding to the  classical
FRW model without the loop geometry modification, there is no
quantum of the volume.

It results from (\ref{eqq5}) that for $b=0$ and $m=0$  the minimum
eigenvalue of $\hat{v}$ equals zero. It is a special case that
corresponds to the classical situation when $v=0$, which due to
(\ref{ham}) means that $p_\phi = 0$ so there is no classical
dynamics (for more details see \cite{Dzierzak:2009ip}). Thus, we
have a direct correspondence between classical and quantum levels
corresponding to this very special state. It is clear that all
other states describe  bouncing {\it dynamics}.

\noindent   {\it 6. Free parameter.} There exists a fundamental
problem underlying LQC (see, \cite{Dzierzak:2008dy} and references
therein), which is the unknown numerical value of the parameter $
\lambda$ \cite{Bojowald:2008ik}.

Determination of $ \lambda$ by standard LQC means
\cite{Ashtekar:2006uz,Ashtekar:2006wn}: (a) considering eigenvalue
problem for the  area operator, $\widehat{Ar} = \widehat{|p|}$, in
kinematical phase space of standard LQC:
$\widehat{Ar}\,|\mu\rangle = \frac{4\pi \gamma l^2_p}{3}\, |\mu|
\,|\mu\rangle =: ar (\mu)\,|\mu\rangle$ so $ar (\mu)$ is
continuous since $\mu\in\dR$; (b) making reference to discrete
eigenvalues, $\{ 0, \Box, \ldots\}$, of kinematical $\widehat{Ar}$
of  LQG, where $\Box := 2\sqrt{3}\,\pi\gamma l^2_p$; and (c) {\it
assuming} that $ ar( \lambda) \equiv \Box$, which leads to $
\lambda = 3\sqrt{3}/2$.

One {\it postulates} in  standard LQC that  a surface cannot be
squeezed to the zero value due to the existence in the Universe of
the {\it quantum} of area.

Physical justification for the assumption on the existence of
quantum of area, offered by standard LQC, seems to be   doubtful
because: (d) $ \widehat{Ar}$ has been examined  in {\it
kinematical} Hilbert space of  LQG, i.e. spectrum of
$\widehat{Ar}$ ignores the algebra of  constraints of LQG so it
has poor {\it physical} meaning; (e)  discrete spectrum of LQG was
used to replace  continuous spectrum of standard LQC, which is the
spectral discretization {\it by hand}; and (f) standard LQC {\it
is not} a cosmological sector of LQG, but a quantization method
{\it inspired} by  LQG \cite{Brunnemann:2005in}.

This is why we propose to treat $\lambda$ as a {\it free}
parameter yet to be determined.

\noindent{\it 7. Conclusions.}  As the Universe expands a discrete
spectrum of the volume operator favors a  foamy structure which
turns into a continuous spacetime with time. The classical FRW
model is commonly used in observational cosmology because it fits
quite well the data. Thus, the detection of any cosmic events
favoring the foamy spacetime would give support to the quantum FRW
model.

Our non-standard LQC method gives results concerning geometrical
properties of space on the {\it physical} phase space so they may
be verified by the data of observational cosmology.

There exist results concerning the spectrum of the volume operator
obtained within LQG (see, e.g.
\cite{Ashtekar:1997fb,Meissner:2005mx}), but cannot be compared
easily with our results due to the lack of a direct correspondence
between LQG and LQC methods (see Sec 6f).

Both standard and non-standard LQC methods offer the resolution of
the initial Big-Bang singularity in the sense that the singularity
is replaced by the regular Big-Bounce (BB) transition. However,
the energy scale specific to BB (the scale of unification of
gravity with quantum physics) has not been determined satisfactory
yet \cite{Malkiewicz:2009zd}. The problem reduces to the problem
of determination of the minimum length \cite{Dzierzak:2008dy}. Can
it be solved by making use of the cosmic data?   There exists
speculation that the foamy structure of spacetime may lead  to the
dependence of the velocity of a photon on its energy. Such
dependance is weak, but may sum up to give a measurable effect in
the case of  photons travelling over cosmological distances across
the Universe \cite{GAC}. Presently, available data suggest that
such {\it dispersion} effects do not occur up to the energy scale
$5\times 10^{17}$ GeV \cite{Aharonian:2008kz}  so this type of
effects may be present, but at higher energies.

The quantum of volume may be used as a measure of a size,
$\lambda_f$, of a spacetime foam. One may speculate that
$\lambda_f := \bigtriangleup^{1/3} = \big(8\pi
G\gamma\hbar\,\lambda\big)^{1/3}$.  Thus, an astro-cosmo data that
determine a size of  spacetime `granularity' $\lambda_f$ may fix
the minimum length parameter $\lambda$ of LQC. That would enable
making an estimate of the critical matter density $\rho_{\max} =
1/2(\kappa \gamma \lambda)^2$ corresponding to the BB
\cite{Malkiewicz:2009qv}.

The granularity of volume should lead to the granularity of energy
of physical fields. We suggest, making use of the de Broglie
relation, that a specific  particle representing a quantum of
energy may have a momentum $p_i$ corresponding to its wavelength
$\lambda_i$ such that $p_i\,\lambda_i = \hbar$. The detection of
an ultrahigh energy particle with specific $p_i$ may be used to
determine $\lambda_i$, and consequently set the upper limit for
the fundamental length $\lambda_f$. The set of parameters
$\lambda_i$ (for a set of particles) may be treated further as
multiplicities of $\lambda_f$ in which case the greatest common
divisor of all $\lambda_i$ would set the lowest upper limit for
$\lambda_f$.

The standard and non-standard LQC methods give comparable results
as they predict the appearance of the Big-Bounce transition
parameterized by a free parameter to be determined
\cite{Malkiewicz:2009zd}. Both methods seem to commute so there
exists an  analogy to the case of quantum electrodynamics
\cite{IBB}. However, our method is fully controlled analytically
as it does not require any numerical work. It may be also linked
with the loop quantum gravity (LQG) by finding relation with the
reduced phase space quantization \cite{Giesel:2007wn}.

\noindent {\it Acknowledgments.} We are grateful to Iwo
Bia{\l}ynicki-Birula, Jan Derezi\'{n}ski, Piotr Dzier\.{z}ak,
Anatol Odzijewicz,  Wies{\l}aw Pusz, and Jakub Rembieli\'{n}ski
for helpful discussions.


\begin{thebibliography}{99}

\bibitem{Heller:2006uq}
  M.~Heller, L.~Pysiak and W.~Sasin,
  Int.\ J.\ Theor.\ Phys.\  {\bf 46}, 2494 (2007).

\bibitem{Rideout:2008df}
  D.~Rideout and P.~Wallden,
  arXiv:0811.1178.

\bibitem{Hamber:2009uz}
  H.~W.~Hamber and R.~M.~Williams,
  arXiv:0907.2652.

\bibitem{Bahr:2009qc}
  B.~Bahr and B.~Dittrich,
  arXiv:0907.4323.

\bibitem{Ambjorn:2009ts}
  J.~Ambjorn, J.~Jurkiewicz and R.~Loll,
  arXiv:0906.3947.

\bibitem{Kaminski:2009fm}
  W.~Kaminski, M.~Kisielowski and J.~Lewandowski,
  arXiv:0909.0939 [gr-qc].


\bibitem{Baez:2009as}
  J.~C.~Baez and A.~Lauda,
  arXiv:0908.2469 [hep-th].

\bibitem{Kostelecky:2008ts}
  V.~A.~Kostelecky and N.~Russell,
  arXiv:0801.0287.

\bibitem{Galaverni:2008yj}
  M.~Galaverni and G.~Sigl,
  Phys.\ Rev.\  D {\bf 78}, 063003 (2008).

\bibitem{AmelinoCamelia:2009pg}
  G.~Amelino-Camelia and L.~Smolin,
  arXiv:0906.3731.

\bibitem{Ellis:2008fc}
  J.~R.~Ellis, N.~Harries, A.~Meregaglia, A.~Rubbia and A.~Sakharov,
  Phys.\ Rev.\  D {\bf 78}, 033013 (2008).

\bibitem{Gleiser:2001rm}
  R.~J.~Gleiser and C.~N.~Kozameh,
  Phys.\ Rev.\  D {\bf 64}, 083007 (2001).

\bibitem{Dzierzak:2008dy}
  P.~Dzierzak, J.~Jezierski, P.~Malkiewicz and W.~Piechocki,
  arXiv:0810.3172.

\bibitem{Malkiewicz:2009zd}
  P.~Malkiewicz and W.~Piechocki,
   Phys.\ Rev.\  D {\bf 80}, 063506 (2009).

\bibitem{Dzierzak:2009ip}
  P.~Dzierzak, P.~Malkiewicz and W.~Piechocki,
  Phys.\ Rev.\  D, accepted, arXiv:0907.3436.

\bibitem{Dzierzak:2009dj}
  P.~Dzierzak and W.~Piechocki,
  arXiv:0909.4211 [gr-qc].

\bibitem{Malkiewicz:2009qv}
  P.~Malkiewicz and W.~Piechocki,
  arXiv:0908.4029 [gr-qc].

\bibitem{Ashtekar:2006uz}
  A.~Ashtekar, T.~Paw{\l}owski and P.~Singh,
  Phys.\ Rev.\  D {\bf 73} (2006) 124038

\bibitem{Ashtekar:2006wn}
  A.~Ashtekar, T.~Paw{\l}owski and P.~Singh,
  Phys.\ Rev.\  D {\bf 74}, 084003 (2006).

\bibitem{Ashtekar:2003hd}
  A.~Ashtekar, M.~Bojowald and J.~Lewandowski,
  Adv.\ Theor.\ Math.\ Phys.\  {\bf 7}, 233 (2003).

\bibitem{Bojowald:2006da}
  M.~Bojowald,
  Living Rev.\ Rel.\  {\bf 8}, 11 (2005).

\bibitem{IBB} I.~Bia{\l}ynicki-Birula, J.~Derezi\'{n}ski and J. Rembieli\'{n}ski,
              private communication.


\bibitem{TT} T. Thiemann \textit{Modern Canonical Quantum General Relativity}
(Cambridge: Cambridge University Press, 2007).

\bibitem{CR} C. Rovelli \textit{Quantum Gravity}
(Cambridge: CUP, 2004).

\bibitem{Ashtekar:2004eh}
  A.~Ashtekar and J.~Lewandowski,
  Class.\ Quant.\ Grav.\  {\bf 21}, R53 (2004).

\bibitem{DS} N. Dunford and J. T. Schwartz, {\it Linear Operators}
(Interscience Publishers, New York, 1958).

\bibitem{RS} M. Reed and B. Simon, {\it Methods of Modern Mathematical
Physics} (Academic Press, San Diego, 1975).

\bibitem{Bojowald:2008ik}
  M.~Bojowald,
  Class.\ Quant.\ Grav.\  {\bf 26} (2009) 075020

\bibitem{Brunnemann:2005in}
  J.~Brunnemann and T.~Thiemann,
  Class.\ Quant.\ Grav.\  {\bf 23}, 1395 (2006)
  [arXiv:gr-qc/0505032].

\bibitem{Ashtekar:1997fb}
  A.~Ashtekar and J.~Lewandowski,
  Adv.\ Theor.\ Math.\ Phys.\  {\bf 1}, 388 (1998).

\bibitem{Meissner:2005mx}
  K.~A.~Meissner,
  Class.\ Quant.\ Grav.\  {\bf 23} (2006) 617

\bibitem{Grain:2009kw}
  J.~Grain and A.~Barrau,
  Phys.\ Rev.\ Lett.\  {\bf 102}, 081301 (2009).

\bibitem{Calcagni:2008ig}
  G.~Calcagni and G.~M.~Hossain,
  arXiv:0810.4330.

\bibitem{Mielczarek:2008pf}
  J.~Mielczarek,
  JCAP {\bf 0811}, 011 (2008).

\bibitem{GAC}
 G.~Amelino-Camelia {\it et al.}, Nature {\bf 393}, 763 (1998).

\bibitem{Aharonian:2008kz}
  F.~Aharonian {\it et al.},
  Phys.\ Rev.\ Lett.\  {\bf 101}, 170402 (2008).


\bibitem{Giesel:2007wn}
  K.~Giesel and T.~Thiemann,
  arXiv:0711.0119 [gr-qc].

\end{thebibliography}
\end{document}